\def\year{2017}\relax
\documentclass[letterpaper]{article} %DO NOT CHANGE THIS
\usepackage{aaai17}  %Required
\usepackage{times}  %Required
\usepackage{helvet}  %Required
\usepackage{courier}  %Required
\usepackage{url}  %Required
\usepackage{graphicx}  %Required
\usepackage{subfigure}
\usepackage{epstopdf}
\usepackage{latexsym}
\usepackage{amssymb}
\usepackage{amsmath}
\usepackage{booktabs}
\begin{document}

\title{Antagonism also Flows through Retweets: \\ The Impact of Out-of-Context Quotes in Opinion Polarization 
Analysis\footnote{ }}
\author{
Pedro Calais Guerra, Roberto C.S.N.P. Souza, Renato M. Assun\c{c}\~{a}o,  Wagner Meira Jr. \\
	     Dept. of Computer Science -- Universidade Federal de Minas Gerais (UFMG)\\
	     \texttt{\{pcalais,nalon,assuncao,meira\}@dcc.ufmg.br}
}

%\author{AAAI Press\\
%Association for the Advancement of Artificial Intelligence\\
%2275 East Bayshore Road, Suite 160\\s
%Palo Alto, California 94303\\
%}
\maketitle
\begin{abstract}
\begin{quote}
In this paper, we study the implications of
the commonplace assumption that most social media studies
make with respect to the nature of message shares (such as retweets) as a
predominantly \emph{positive} interaction. By analyzing two large longitudinal Brazilian Twitter
datasets containing 5 years of conversations on two polarizing topics -- Politics and Sports,
we empirically demonstrate that groups holding antagonistic views can actually
retweet each other \emph{more often} than they retweet other
groups. We show that assuming
retweets as endorsement interactions can lead to misleading conclusions with respect to the level of antagonism
among social communities, and that this apparent paradox
is explained in part by the use of retweets to quote
the original content creator out of the message's original temporal context,
for humor and criticism purposes. As a consequence, messages diffused 
on online media can have their polarity \emph{reversed} over time, 
what poses challenges for social and computer
scientists aiming to classify and track opinion groups 
on online media. On the other hand, we found that the time users take to retweet a message after it has been originally posted can be a useful signal to infer antagonism in social platforms, and that surges of out-of-context retweets correlate with sentiment drifts triggered by real-world events. We also discuss how such evidences can be embedded in sentiment analysis models.

\end{quote}
\end{abstract}

\section{Introduction}
\label{sec:introduction}
In this paper, we study the implications of
the commonplace assumption that most social media studies
make with respect to the nature of message shares (such as retweets) as a
predominantly \emph{positive} interaction. Given that on general purpose social platforms such as Facebook and Twitter there are 
no explicit positive and negative signs encoded
in the edges, it is commonly assumed (in general, implicitly) that a connection among users through
message shares indicate increased homophily among them~\cite{pcalaisKDD11,PoliticalPolarizationTwitter}.
In general, studies of polarized online communities induced by topics such as Politics and public
policies do not conduct any explicit analysis of antagonism at the edge granularity, and
the degree of separation between communities as well as the controversial nature of the topic is accepted as sufficient evidence of polarization~\cite{QuantifyingControversySocialMedia}. We provide a qualitative and quantitative analysis on the use of retweets
as \emph{negative} interactions. In particular, we analyze two large Brazilian Twitter datasets
on polarizing topics -- Politics and Soccer -- which lead us to four main findings related to behavioral patterns on social-media based interactions:

\begin{enumerate}

\item Antagonistic communities tend to share each other's content \emph{more often} than they
share content from other less polarizing and conflicting groups. The immediate consequence of this
observation is that a simplistic consideration of retweets as an endorsement interaction can lead to misleading
conclusions with respect to the nature and polarity of group relationships, as
a large number of retweets flowing from one community to another may be misinterpreted
as a signal of support. 

\item We observe retweets employed
as a mechanism for \emph{quoting out of context}, 
a known strategy of reproducing a passage or quote out of its original
context with the intent of distorting its intended meaning~\cite{QuotingOutOfContext}.
In particular, we found that Twitter users share old messages
posted by someone from an opposing side 
 with the goal of creating irony when putting
the message out of its original temporal context. We observed that some messages
are broadcasted even 6 years after they have been originally posted, with the intention of reinforcing an antagonistic and contrary position, rather than indicating support. In our datasets, a significant fraction of retweets crossing antagonistic communities
are out of context retweets.

\item As a consequence of Finding 2, messages diffused in a social platform can actually have their polarity \emph{reversed} over time,
since the first users sharing the message endorse its original intended content, while other users share
the message in response to a real-world event aiming to satirize and to prove that the message's author was wrong, attaching to it
an implicit negative polarity. This concept drift poses interesting challenges for research in text-based sentiment analysis and sarcasm detection.

\item Real-world events can trigger a burst of such out-of-context retweets. We show how the distribution of retweet response
times in a concentrated time span can be a signal which helps detecting sudden sentiment drifts among opinion groups, as they focus
on retweeting old tweets from their adversaries during specific real-world events.

\end{enumerate}

We believe the main reason these findings
on the use of retweets to convey disagreement
 remain unnoticed in
the social network analysis literature is due the focus on research on bipolarized social networks, characterized by the emergence of \emph{exactly} two dominant conflicting groups, such as republicans versus democrats~\cite{DividedTheyBlog}, pro and anti gun-control~\cite{pcalaisICWSM13},
and pro-life versus pro-choice voices. In this setting, once you determine (automatically or by manual examination) the leaning of a group
toward a controversial topic, their (negative) opinion w.r.t.\@ the opposite viewpoint is implicitly determined, and 
no further analysis of edge polarities is usually performed.

\begin{figure}[!htb]
\centering
\subfigure[2014 Brazilian Political Twitter.]{\label{fig:BrazilPolitics}\includegraphics[width=0.28\textwidth]{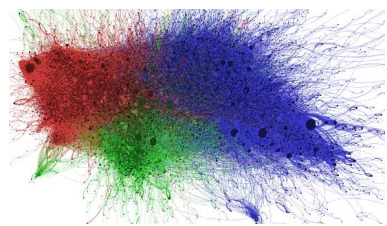}}
\subfigure[2010-2016 Brazilian Soccer debate in Twitter.]{\label{fig:brazilian_soccer}\includegraphics[width=0.30\textwidth]{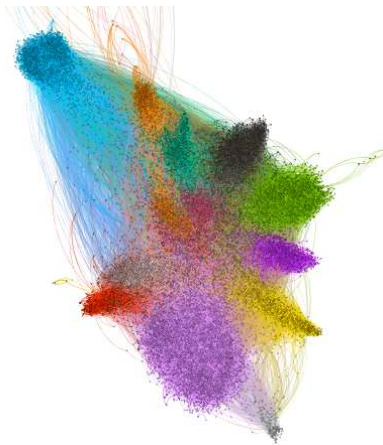}}
\caption{On the top, a network of retweets obtained from Twitter showing
3 communities formed around the 3 main candidates in the 2014 Brazilian
Presidential Elections. On the bottom, communities formed around the 12 top Brazilian Soccer teams. Although both topics
are polarizing in nature, in a \emph{multipolarized} domain not every pair of groups is expected to share antagonism.}
\label{fig:multipolarized_graphs}
\end{figure}

To remove the straight-forward polarity assignment of bipolarized communities and analyze the interplay
between retweets and (lack of) antagonism, we collected datasets on discussion domains where more than two communities interact,
namely, political discussion in a multipartisan political system and multiple groups of sports fans engaging on conversations about the Brazilian Soccer League. In Figure~\ref{fig:BrazilPolitics}, we plot in different colors the three largest communities
found in a network of retweets we collected from Twitter during the 2014 Brazilian Presidential Elections, representing
groups of people formed around the 3 main candidates (Dilma Rousseff, A\'ecio Neves and Marina Silva); in Figure~\ref{fig:brazilian_soccer}
we do the same for the 12 largest exchanging messages about Brazilian soccer. Differently 
from bipolarized social graphs, since now there are $K>2$ possible sides one
user may belong to, the identification of an individual as a member of a community does not necessarily imply on antagonism
with respect to all the remaining $K-1$ groups; each group member can be indifferent, or neutral, to a subset of the remaining groups, or even
support more than one group simultaneously. As a consequence, we need to conduct a deeper analysis of retweets crossing
communities to gain insights on group relationships.

Our work contributes to social media research in
two distinct directions. Findings 1 and 3 add to the recent trend on the
pitfalls and drawbacks of making inferences based on social
media data~\cite{Snowden,ChallengesSocialMedia,HowNotToPredictElections}.
Findings 2 and 4, on the other hand, explore how temporal information associated
to retweets can be a rich signal to be incorporated
into models focused on antagonism detection and real-time tracking of opinions in social media.

In the remainder of this paper, we first discuss related work
on polarization and unsigned edges in social networks. Next we analyze two longitudinal
Twitter datasets to empirically
demonstrate that, on multipolarized social networks,
assuming retweets as positive interactions can be misleading. Finally, we characterize how
cross-group retweets differ from intra-group retweets with respect to the
distribution of the time differences between the message posting time and the retweet action, and
we show how this signal can be embedded into models 
that aim to detect the controversy level among
opinion groups and real-time sudden drifts on their sentiment and opinions.

\section{Related Work}
\label{sec:related_work}
\label{sec:related_work_multi}

 On social networks whose edge signs are labeled, antagonistic relationships among
communities are naturally reflected by the number of positive and negative edges flowing 
from the source community to a target community, and the communities themselves can
be found by algorithms especially designed to deal
with negative edges~\cite{SpectralAnalysisSignedGraphs,BayesianApproachSignedNetworks,MiningAntagonisticCommunitiesExplicitTrust}.

Many works qualitatively discuss and document the empirical observation that unlabeled social interactions
on general purpose social platforms such as Twitter and Facebook can convey negative sentiment: replies and comments, as web hyperlinks, do not carry an explicit sentiment
label and can be either positive or negative~\cite{PredictingPositiveNegativeLinks,PredictingPositiveNegativeLinksTransferLearning}.
Message broadcasts, on the other hand, have been categorized by early works on behavioral analysis on Twitter
as a strictly positive interaction~\cite{TweetTweetRetweet}. As users expertise evolved, they had begun finding uses of retweets
that do not convey agreement. ``Retweets are not endorsements'' is a common
disclaimer found in biographies of journalists and think tankers in Twitter, whereas some people share stuff that they vehemently disagree only to show the idiocy of the people they oppose. One can also broadcast the original message and append comments to it (``quote RTs'', in Twitter), often in disagreement with the original content, what also contributes to turn shares and retweets into an ambiguous signal with respect to the sentiment they convey~\cite{QuoteRTs}. In summary, retweets and shares are often a ``hate-linking'' strategy -- linking to disagree and criticize, often in an ironic and sarcastic manner, rather than to endorse~\cite{BigQuestionsBigData}.

Although documented in the literature as a known behavior, the impact of such ``negative" retweets
on community and network analysis has not been 
the focus of in-depth studies so far. Usually, social network analysis practitioners assume, implicitly or explicitly, that retweets (or more generally, shares) have a predominant endorsement nature. A
recurrent pattern in community analysis works making sense of social media datasets is that they limit their analysis to
social networks whose dominant topic induces a partition of the graph into exactly two conflicting sides: liberal versus conservative parties, pro-gun and anti-gun voices, pro-choice and pro-life~\cite{PoliticalPolarizationTwitter,PartyOverHere,DividedTheyBlog,QuantifyingPoliticalLeaningFromTweetsRetweets}. As we will
show in the next sections, in bipolarized scenarios, it is harder to grasp the use of retweets to convey disagreement.

Our contribution in this paper is twofold. While we raise awareness to the network science community of the
implications of assuming retweets as positive interactions,
we propose a new edge-level signal -- the retweet response time, i.e. the
amount of time the user took to hit the retweet
button after the original message has been posted -- 
to help disambiguating positive from
negative edges in a social network containing timestamped edges.

\section{Data Collection and Preparation}
\label{sec:datasets}
We used Twitter's Streaming API\footnote{Twitter Streaming API: \\ \url{https://dev.twitter.com/streaming/overview}.} to monitor two topics that motivate intense debate
on offline and online media and thus are suitable for analysis
of formation of antagonistic communities:
Politics~\cite{pcalaisKDD11} and Sports~\cite{LiveSportsEvents}. Table~\ref{tab:dataset_summary}
provides details on the datasets. 

\begin{table}[!htb]
  \centering
		  \caption{General description of the two Twitter datasets we consider. Note the
	large variability on (native) retweet response times. }
  \begin{tabular}{r|c|c}
    \toprule
    & \multicolumn{2}{c}{\small{\textbf{Topic}}} \\ \hline
    {\small\textbf{}}
    & {\small \textit{Politics}}
    & {\small \textit{Soccer}} \\
    \midrule
    period & 2010-16 & 2010-16 \\
		\# groups & 3 & 12 \\
    \# tweets & 20.5 M & 103M \\
		\# users & 3.1M & 8.7M \\ \hline
		manual RTs & 46K & 2K \\
		quote RTs & 67K & 3K \\
		native RTs & 9.1M  & 30.9M \\ \hline
		RT mean response time (hours) & 29.5h  & 43.5h  \\
		RT median response time (hours) & 0.24h  & 0.23h  \\
		RT response time std (hours) & 255.4h  & 368.7h  \\ \hline
		\# replies & 3.2M & 20.8 M \\
		reply mean reaction time (hours)  & 5.1h & 3.5h \\
    reply reaction time  std (hours) & 188.3h & 194.0h \\ 
		\bottomrule
  \end{tabular}

	~\label{tab:dataset_summary}
\end{table}

In the political topic, our data collection was driven by the main candidates in the
2010 and 2014 Brazilian presidential elections, including Dilma
Rousseff, elected for the presidency in both years. In December 2015, Ms.\@ Rousseff faced an impeachment trial conducted
by the Brazilian Congress, and on May 12th, 2016, the Senate voted to suspend her for 180 days.
The vice-president Michel Temer, elected with her in 2010 and 2014, assumed as the provisory president. We
monitored mentions to politician Twitter profiles and names, the hashtags used by each side participating in the political debate and the names of the presidents of the Brazilian
Lower House and the Senate, which directly conducted Ms. Rousseff's impeachment process in the Congress.

We also collected public tweets about the 2010 to 2016 editions
of the Brazilian Soccer League. We monitored mentions to the 12 largest Brazilian soccer teams and match-related
keywords, such as ``goal'', ``penalty'' and ``yellow card''. 

Notice that the fact that we collected tweets during
a time span of more than five years allow us to extract the time interval between the original message and each of its retweets, and
observe large deltas between these timestamps. We call
this time interval the \emph{retweet response time}. Table~\ref{tab:dataset_summary} shows that the
mean retweet response time is in the magnitude of several hours and
it is an order of magnitude higher than the median retweet response time.
Also, its standard deviation is almost an order of magnitude larger
than the mean, what indicates a high variability in retweet response times.
 Compared to replies, the average response time of retweets is about 6 and 12 times higher,
in the Politics and Soccer dataset, respectively. This suggests
that there might be some specific behavioral and temporal patterns associated with retweets.
We will show how such `late retweets' relate to polarization and interactions
among antagonistic groups later in this paper.

Three types of retweets can be extracted from the raw JSON tuples: manual retweets, i.e.,
messages manually created in the format `RT @username message'; a quote retweet, when the user prepends or appends a comment
to the original message (as in `Cool! RT @username message'); and a retweet triggered through the native Twitter
retweet button. We have chosen to focus our analysis on native retweets
for three reasons:

\begin{enumerate}
\item They represent the vast majority of retweets (see Table 1);
\item Although manual and quote retweets are also legitimate user interactions, native retweets 
better reflect how the user interface design affects user behavior, 
since they are directly implemented in Twitter's user interface;
\item In a native retweet, the original tweet posting time is provided in the JSON format; therefore we do not need
to have collected the original message in order to compute the retweet response time. 
\end{enumerate}

\textbf{Community detection.} Once collected we prepared the data for our various analysis as described next.

The first step is to partition the social network induced by the messages and
represented as a graph $G(V,E)$ into meaningful communities. 
Although our methodology does not depend on the specific graph clustering algorithm,
finding communities on polarized topics is eased by the fact that it is usually
simple to find seeds -- users that are previously
known to belong to a specific community. In the case of the Twitter datasets we take into consideration, the official
profiles of politicians,  political parties and soccer clubs
are natural seeds that can be fed to a semi-supervised clustering
algorithm that expands the seeds to the communities formed around them~\cite{pcalaisKDD11,Snowden,KleinbergSmallSeedSets}. 

\begin{figure}[!htb]
  \centering
  \includegraphics[width=0.35\textwidth]{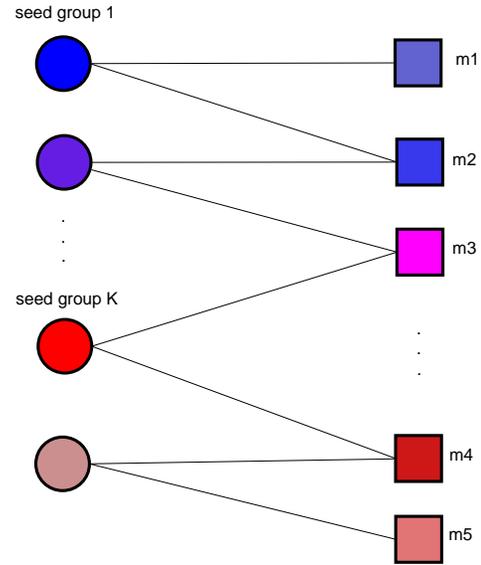}
  \caption{A bipartite user-message graph connecting users with messages they interact with. To find communities, we run a random walk with restarts from each seed that represents a community (notice in the figure that they are explicitly labeled); the random walker will traverse more frequently the links and nodes belonging to the community the seed belongs to. Node colors represent relative proximities to the
	the red/blue sides.}
  \label{fig:example_bipartite_graph}
\end{figure}

Different graphs can be built based on the datasets described in Table 1;
traditionally, a social network $G(V,E)$ represents a set of users $V$ and a set of edges $E$ that connect two users if they exceed a threshold of interaction activity. The limitation of this modeling is that it hides the individual user-message interactions: for instance, two users holding opposite opinions may propagate different messages from the same media outlet, what could wrongly indicate that both share the same opinion. Connecting users directly hides the fact that the individual messages may have a potentially
different sentiment with respect to different entities; i.e., a media outlet may post a positive message w.r.t to a politician one day and a negative message a week later. By representing interactions in a user-message bipartite retweet graph, as shown in Figure~\ref{fig:example_bipartite_graph}, we keep this more granular information.

We assume that the number of communities $K$ formed around a topic $T$ is known in advance and it
is a parameter of our method. To estimate user and message leanings toward each of the $K$ groups,  we employ a label propagation-like strategy based on random walk with restarts~\cite{RWR}: a random walker
departs from each seed and travels in the user-message retweet bipartite graph by randomly choosing an edge to decide which node it should go next.
With a probability (1 - $\alpha) = 0.85$, the random walker restarts the random walking process from its original seed. As a consequence,
the random walker tends to spend more time inside the cluster its seed belongs to~\cite{pcalaisKDD11}.
Each node is then assigned to its closest seed (i.e., community), as shown
in the node colors in the toy example from Figure~\ref{fig:example_bipartite_graph}. For more details on the random walk-based community detection algorithm,
please refer to~\cite{pcalaisKDD11}.

For both the Politics and Soccer Twitter dataset, we performed a validation of the $K$ communities we found using a sampling strategy on the correlation between communities and profiles that make explicit their side. In particular, we exploit the evidence provided by many Twitter users that append to their profile names the soccer  team or political party they support; and, in general, the content they publish will favor the respective mentioned side, as we observed through manual inspection of a sample. For example, @[first name][name of favorite soccer team] is a common account name pattern, as in @JohnCruzeiro,  through which John declares he is a Cruzeiro fan. From the 13,892 messages these specific users generated in the Elections and Soccer dataset, we found that 91.53\% were assigned by our algorithm to the community indicated in their profile name. Although we acknowledge that user account names are an imperfect ground truth and these users are more likely to present a more active and clearly defined behavior and thus they are more easily classified, we believe this number indicates that the accuracy of the random-walk based clustering method is enough for our data analysis purposes.

\section{Finding 1: antagonistic groups retweet each other more than they retweet other groups}
\label{sec:paradox}
\label{sec:multi_paradox}

As we pointed out in Section 1, the polarity relationships among
the $K$ communities found is not an explicit byproduct of a community detection method whose input is an unsigned graph. Recall that, on bipolarized domains, no subsequent analysis is usually performed, other than the quantification of the
degree of separation between the pair of communities, using community quality metrics such as \emph{modularity}~\cite{PartyOverHere,DividedTheyBlog}.
It is a standard practice to assume that the more separated the communities are, the more antagonism is observed, as a consequence
of the homophily principle~\cite{BirdsOfAFeather}. 

The intrinsic limitation of a bipolarized network is that only
one separation metric value can be computed, since there is
only one pair of communities. Since we are studying $K > 2$
cases, we now have $K \choose 2$ pairwise
community metrics to compare. For the sake of simplicity, for each
pair of communities we compute the proportion of retweets triggered from users belonging to community $i$ that flow toward messages posted by members of community $j$
relative to all retweets that community $i$ trigger to the other groups in the graph:

\begin{equation}
RT\_ratio(i,j) = \frac{RT_{i,j}}{\sum\limits_{k=1, k \ne i}^{K}RT_{i,k}}
\end{equation}

We compare $RT\_ratio(i,j)$ considering 
the known local rivalries that exist in Brazilian Soccer among
soccer clubs from the same Brazilian state, as listed in Table~\ref{tab:local_rivalries}.

\begin{table}[h]
  \centering
	      \caption{Local rivalries in Brazilian Soccer. Stronger antagonism exists between soccer clubs and communities of supporters belonging
			to the same Brazilian state.}
\begin{tabular}{c|c} \hline
  \textbf{Brazilian state}  & \textbf{local rivalries} \\\hline
  Minas Gerais &  Cruzeiro, Atl\'etico \\
	S\~{a}o Paulo &  SPFC, Santos, Corinthians, Palmeiras \\
	Rio G. do Sul & Gr\^{e}mio, Internacional \\
  Rio de Janeiro &  Flamengo, Fluminense, Vasco, Botafogo \\
			\bottomrule
\end{tabular}

\label{tab:local_rivalries}
\end{table}

\begin{figure}[!htb]
\centering
\includegraphics[width=0.45\textwidth]{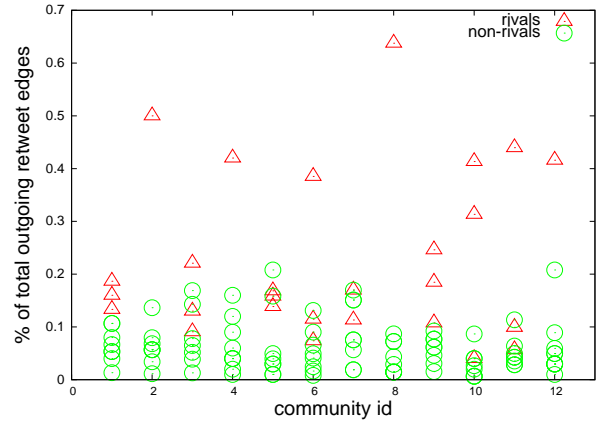}
\caption{$RT\_ratio(i,j)$ for each pair of 12 communities discussing Brazilian soccer in Twitter. More antagonistic communities retweet each other more than neutral, less polarizing communities.}
\label{fig:SoccerParadox}
\end{figure}

In Figure~\ref{fig:SoccerParadox} we plot $RT\_ratio(i,j)$
for all the $K \choose 2$ pairs of communities formed around supporters of Brazilian soccer clubs,
and we visually discriminate between pairs of rival communities (red triangles)
and non-rival communities (green circles) according to the ground truth from Table~\ref{tab:local_rivalries}. The graph shows
a somewhat unexpected result: pairs of communities that are \emph{more antagonistic}
 (i.e., the opposing sides belong to the same Brazilian state) tend to retweet each other's content \emph{more often} than
when there is less, or no antagonism between them. For
example, Cruzeiro's community (id = 8) targets about
65\% of its cross-group retweets to Atl\'etico's community, their
sole fierce rival in Brazilian state of Minas Gerais. As another example, community
1, which identifies supporters from Rio de Janeiro team Flamengo, prefers
to retweet messages for their three local rivals. As a general rule, red triangles dominate green circles,
i.e., retweets are targeted more often to antagonistic communities than to more neutral, less conflicting
groups.

The fundamental insight to learn from Figure~\ref{fig:SoccerParadox}
is that retweets carrying a negative polarity directly impact
the network structure and make antagonistic communities \emph{closer} in
the social graph. On traditional bipolarized
domains in which current literature focuses, this apparent
paradox is inherently unnoticeable, since there is only a single pair of
antagonistic communities and thus only a single separation
metric to be computed. 

We list a few intents that motivate Twitter users in
retweeting messages they disagree with:

\begin{itemize}

\item \textbf{Share to show contrary opinion.} Many times,
a user propagates a message he or she disagrees with to
show the message to their followers or friends and comment
on that content. The goal is to start a discussion and gauge reactions.

\item \textbf{Fake or edited retweets.}
We do not include these retweets in our
analysis, but some Twitter users create fake retweets, in the
format \emph{``RT @user fake message''}, assigning
to \emph{@user} a message that has never has been posted. Fake
retweets have already being investigated as a spamming
activity in Twitter~\cite{TwitteringMachine}, in which
spammers try to borrow from the reputation of celebrities. In
the context of polarized discussions, however, the goal is different -- 
to make criticism or even spread false information~\cite{EditedRetweets}.

\item \textbf{Out-of-context quoting.} We will provide
an in-depth analysis of this behavior in the next section.
In summary, a user propagates a message he or
she disagrees with and puts it out of context,  in
order to create sarcasm or irony. In this case, we usually
see messages being shared long after they
were originally posted, typically when 
the original message stated a prediction that turned
out to be false later.

\end{itemize}

\textbf{Negative retweets and the filter bubble.} In a recent study
by Pew Research Center, polarized discussions have been identified as one of the top 6
most common conversational structures in Twitter~\cite{PewPolarization}. For that reason,
better understanding the social structures induced by polarized debate is important because polarization of opinions induces segregation in the society, causing people with different viewpoints to become isolated in islands where everyone thinks like them~\cite{BiasTrust}. Such filter bubble caused by social media
systems limits the exposure of users to ideologically diverse content, and is a growing concern~\cite{DoesFacebookIntroduceIdeologicalBias,ExposureIdeologicallyDiverseNews}. The behavioral pattern we 
document here has the unintentional side effect of \emph{reducing} the filter bubble, letting followers
of advocates of one viewpoint to get to know the opinions of the other side.

The ``paradox'' of antagonistic communities being linked by
more retweets make clear some assumptions which are commonly implicitly made in the literature
with respect to the treatment of edge signs. While the correctness and applicability of each assumption
depends on inherent characteristics of each dataset, we advocate that it is a good practice
to make it clear the expectations with respect to the following aspects/metrics:

\begin{enumerate}

\item \textbf{Edge sign prior}. The
vast majority of community detection methods on social media networks
are built over the assumption -- which is, most of the time,
not make explicit -- that there is an apriori knowledge that edges are more likely to be positive than negative.
If $P(sign(edge) = +)$ is sufficiently high, it
is reasonable to expect that the method will output the identification of groups of users and messages around
a cohesive viewpoint and high level of homophily. For instance, in a blog citation network, one blog may cite the other to disagree with it, but since most of the time a blog citation is an endorsement rather than a disapproval, edge label-agnostic
community detection methods work reasonably.

\item \textbf{Antagonism and community separation metrics.} It is
a standard practice to measure the degree of antagonism between communities
through separation metrics such as modularity, considering that the more separated the communities are, the higher their level of antagonism
and controversy~\cite{DividedTheyBlog,PoliticalPolarizationTwitter}. However, a smaller modularity may actually indicate an increase of antagonism through interaction via negative retweets and debate through replies and comments.

\item \textbf{Domain of discussion and antagonism}.
The other implicit assumption usually made by social network analysis researches on networks
subject to polarization is that the domain implicitly denotes antagonism, rather than being inferred
from a principled method that analyzes the network structure and content. More formally,
it can be assumed that, once you condition on edges that cross communities, the likelihood
of an edge being negative is now greater than being positive. As a consequence, once users are grouped into two communities, members of one group will automatically be assigned to have a contrary or antagonistic opinion regarding the remaining group. These
works do not deal with differences between antagonism or indifference, neither with a more accurate handling
of edge signs.
\end{enumerate}

In the next section, we will use the temporal context where
retweets occur as evidence that indicates which retweets have a higher probability
of conveying antagonism.

\section{Finding 2: out-of-context retweets are more prevalent on cross-group relationships}
\label{sec:reaction_times}

We are now interested in understanding 
differences between internal retweets, i.e.,
those which connect users and messages
belonging to a single community, and cross-group
retweets, i.e., those which are triggered by
users from one community but propagate a message
posted by an user from another group.

We focus our analysis on the retweet response time --
the time interval between the original message
posting time and the retweet time. Previous
studies found that 50\% of retweets tend to occur up to one
hour after the original message posting time~\cite{WhatIsTwitter};
other studies have related very short and very long retweet response times to fraudulent
activity to boost user popularity~\cite{FraudRetweet}. Our goal is to analyze
retweet response time under the perspective of the message polarity
and the polarity that the user broadcasting the message is attempting to convey.

\begin{figure}[!htb]
\centering
\includegraphics[width=0.45\textwidth]{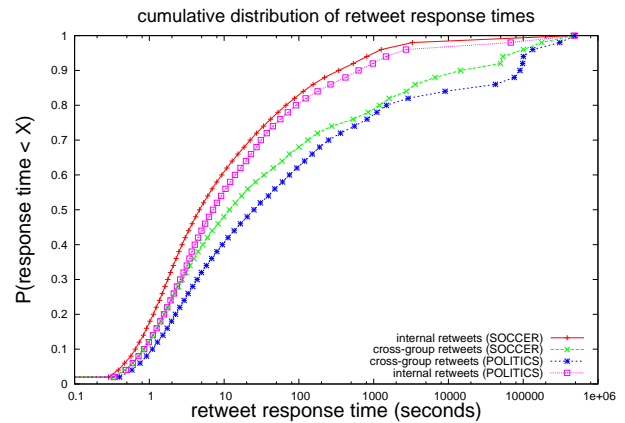}
\caption{On average, retweets which cross antagonistic communities tend have larger response times
than inter-community retweets. This empirical observation suggest the potential use of retweet response
times as a qualifying signal for prediction of edge labels and community memberships.}
\label{fig:CDF_ReactionTime}
\end{figure}

In Figure~\ref{fig:CDF_ReactionTime} we plot the cumulative distribution of retweet response times, measured in seconds. We plot
this distribution for internal (intra-community) and cross-group (inter-community) retweets for both the Soccer and Politics dataset. Notice
that cross-group retweets tend to occur later when compared to internal
retweets. For instance, at least 30\% of retweets
connecting groups in both datasets occur after 16 hours of the original message posting time; on the other hand,
in the case of internal retweets, only 10\% of retweets occur temporally far from
the original post. Notice, also, that the four curves group into two clusters, indicating
that in both topics the retweet response time distribution is similar.

\begin{table*}[htp!]
  \centering
      \caption{The top 5 most retweeted messages from Brazilian VP (@MichelTemer) during impeachment voting period were very old retweets.
			  Users retweeted old messages indicating support from Temer to Dilma, although the moment was of tension and conflict between them.
			 }
\begin{tabular}{c|c|c} \hline
  \textbf{tweet}  & \textbf{\# retweets} & \textbf{avg. retweet response time (days)} \\\hline
  \emph{``We will shout loud to everyone: ``Dilma is our President''''.} & 9,669  & \textbf{606} \\
	\emph{``Impeachment is unthinkable and has no basis in law neither in Politics.''} & 9,338  & \textbf{385} \\
	\emph{``Dilma is the best person to conduct our country.''} & 5,031 & \textbf{628} \\
	\emph{``Congratulations on your birthday, Dilma. God Bless You.''} & 2,020 & \textbf{857} \\
	\emph{``Dilma is displaying confidence and knowledge.''}  & 1,627 & \textbf{2105} \\ \hline
%			\bottomrule
\end{tabular}

% \vspace {-5pt}
\label{tab:top5_michel}
\end{table*}

\begin{table*}[htp!]
  \centering
      \caption{2 of the top 5 most retweeted tweets from Brazilian President (@dilmabr) during impeachment voting period were very old retweets, indicating support from Dilma to Temer. Dilma, however, were accusing her VP to plan a coup against her.}
\begin{tabular}{c|c|c} \hline
  \textbf{tweet}  & \textbf{\# retweets} & \textbf{avg. retweet response time (days)} \\\hline
  \emph{``I thank my VP Michel Temer for all the support.''} & 4,314 & \textbf{538}  \\
  \emph{``The impeachment is against the wishes of the Brazilian people.''} &  3,635 & 1.21  \\
  \emph{``Follow President Dilma live from Periscope.''} & 684  & 0.58  \\
	\emph{``President Dilma will make a speech on the Brazilian Senate decision.''} & 606  & 0.39  \\
	\emph{``Our VP @MichelTemer is now on Twitter. Let's welcome him!''} & 329 & \textbf{693}  \\	 \hline
	
%			\bottomrule
\end{tabular}

% \vspace {-5pt}
\label{tab:top5_dilma}
\end{table*}

We now take a closer look at some messages. For instance, consider the following tweet posted by the official account of the Brazilian elected
vice-president Michel Temer about a speech given on TV by his presidential candidate, Dilma
Rousseff, during the 2010 Presidential Elections:

\begin{quote}
2010-08-05 11:11 PM: @MichelTemer: \emph{Dilma is displaying confidence and knowledge.}
\end{quote}

Six years after this post, President Rousseff has been suspended by
the Brazilian Congress following an impeachment trial
of misuse of public money. In response, she gave
a speech on March 12th, 2015 accusing VP Temer's party (PMDB)
to plan a coup against her. During her speech, many users contrary
to Rousseff began retweeting Temer's 2010 message:

\begin{quote}
2016-05-12 12:23 AM: @randomRousseffOppositor: RT @MichelTemer: \emph{Dilma is displaying confidence and knowledge.}
\end{quote}

This is a clear attempt to retweet a message with the intention to attach to it a negative connotation; it does not support
nor endorse its original content. On the contrary, retweeterers of this message in 2016 attach to it a semantics which is exactly
the opposite to the one stated in the direct interpretation of the message, what is precisely the definition of irony~\cite{ComputationalIrony}. While the ``contextomy'' practice usually refers to selecting specific words from their original linguistic context~\cite{QuotingOutOfContext}, we
see that, in Twitter, such change of meaning is usually associated with some temporal evolution.

Politicians are often targeted by out of context quotes~\cite{TheyNeverSaidIt}. Tables~\ref{tab:top5_michel} and~\ref{tab:top5_dilma}
list the most popular tweets from @MichelTemer and @dilmabr which
received retweets during the impeachment voting process period. In case of VP Temer, all top 5 
most retweeted tweets are very old tweets; and the same applies to 2 of the top 5 messages
from Dilma Rousseff. All those messages indicate affective and positive relationships 
among both politicians, even though the moment was of conflict between them due to
the impeachment trial. As a consequence, content-based and network-based algorithms
built over the retweet-as-endorsement assumption can easily be led to make wrong
predictions over this data.

\subsection{Late retweets and Twitter user attributes}

To further explore how retweet response times can be an explanatory signal that helps on
various social-related prediction tasks, we investigate how late
retweets are disproportionately targeted to some types of Twitter users. In particular,
we calculated the prevalence of late retweets targeting messages posted by three types of users:

\begin{enumerate}
\item Verified users; i.e., users who own a blue verified badge
assigned by Twitter to let people know that an account of public interest is authentic.
In the Politics dataset, only 17\% of the retweets target verified users.

\item Users who have a large follower base; we classified in this category users who have at least 100,000 followers.
In the Politics dataset, 23\% of retweets target such users.
\item Users who have been retweeted by users who were also retweeted by them. In the Political dataset, only 2\% of retweets are triggered by
reciprocal retweeterers.

\end{enumerate}

\begin{figure}[!htb]
\centering
\includegraphics[width=0.45\textwidth]{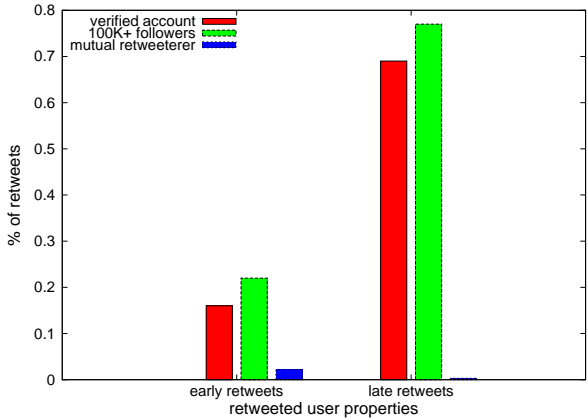}
\caption{Late retweets are disproportionately targeted to users owning verified accounts and a large follower base. 67\% and 77\% of late retweets target verified and large-follower based users, respectively. On the other hand, recripocal
retweeterers are less often involved in late retweets. Results are similar in the Soccer dataset.}
\label{fig:histogram_accounts}
\end{figure}

For the sake of this analysis, we considered a retweet
to be ``late'' if its response time is at least two
standard deviations greater than the average response time.
 In Figure~\ref{fig:histogram_accounts}, we observe that,
when compared to ``early'' retweets, late retweets
disproportionately target messages from verified users, and users
who have a large follower base. In both cases, more than two
thirds of late retweets target those types of users. Furthermore,
we see that users who mutually retweet each other are less
likely to be targeted by a late retweet.

Those measures reinforce a few hypotheses. The first is that late retweets
are most commonly targeted to famous and well-known users because they
provide \emph{context} to support the ironic and sarcastic purpose of retweeting
their tweets out of their original temporal context.
Second, the observation that mutually-retweeted users are less likely to be involved
in a late retweet is an indication that late retweets tend
to be negative interactions, since reciprocal interactions
have been shown to be correlated to homophilic ties~\cite{TwitterRank,WhatIsTwitter}.

 While in isolation it is hard to tell whether a retweet is an endorsement or not, new
signals captured from the social and temporal context, such as the retweet response time, can help on
the design of community detection methods. Negative retweets also pose challenges to signed network analysis: as we showed, the sign of an edge in the social graph may actually depend on \emph{when} the edge has been created, what suggests that
embedding temporal information on edge creation information may enhance signed network models and algorithms that
focus on prediction tasks such as social tie predictions.

\section{Finding 3: Message polarities may reverse over time}

One implication of out-of-context retweets is that messages' polarities
can actually \emph{reverse} over time. Consider this tweet posted by a popular
profile representing the Brazilian soccer club Atl\'{e}tico Mineiro posted in early 2013
mentioning their rivals Cruzeiro:

\begin{quote}
2013-02-02  10:20 PM @caatleticomg: \emph{2013 will be a great year for Cruzeiro: financial debt and injured players.}
\end{quote}

\begin{figure}[!htb]
\centering
\includegraphics[width=0.43\textwidth]{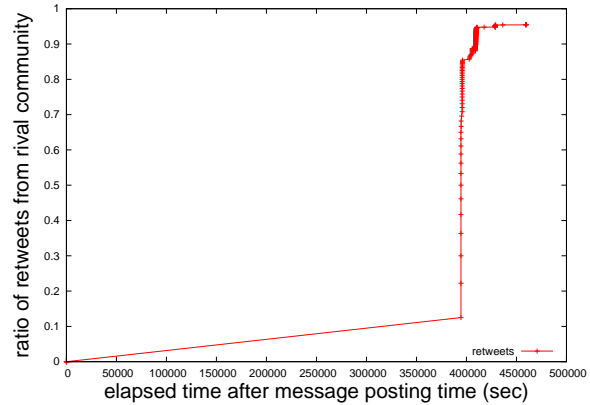}
\caption{Message polarity may reverse over time: a message initially negative to Cruzeiro's supporters has turned into positive, after
being retweeted with an ironic intention after Atl\'{e}tico fans predictions on Cruzeiro performance have failed.}
\label{fig:msg_polarity_switching}
\end{figure}

``Great'', here, was employed in an ironic way: the tweet was actually predicting (and wishing) a bad year for its rival Cruzeiro. At that year, however, Cruzeiro enjoyed one of the best league performances of its history, winning the national league after scoring 76 points, eleven more than the runner-up. In Figure~\ref{fig:msg_polarity_switching}, we show the proportion of retweets of this message originating from Cruzeiro's supporters over time; the original message was posted at time 0. Notice that 400,000 seconds (277 days) after the message
has been originally posted, there is a sudden drift in the ratio of retweets originating from Cruzeiro's supporters;
it goes from a negligible ratio to about 95\% of retweets. The change on the dominant group retweeting the message happened when Cruzeiro won the Brazilian National League
and fans were celebrating, and they wanted to make clear that the ironic prediction from its rivals
have flagrantly failed. 

Since the same content can convey an opposite sentiment depending of its temporal context,
text-only irony and sarcasm classifiers such as~\cite{WordEmbeddingsSarcasm} will
not be able to correctly predict the intent of the message propagator. In fact,
context plays a significant role on human communication~\cite{ComputationalIrony} and the polarity reversal we witness here calls
for more context-aware signals on sarcasm detectors, what includes temporal and social features in models.

\section{Finding 4: Spikes of late retweets correlate with sentiment drifts}

In Section 5, we showed that out-of-context retweets
have an increased chance of being a negative interaction. We now
investigate whether there is a concentration of such retweets in 
specific time frames. We focus on the Soccer topic, more specifically,
in the year of 2013, which was particularly 
eventful for Atletico and Cruzeiro supporters.

\begin{figure}[!htb]
\centering
\includegraphics[width=0.45\textwidth]{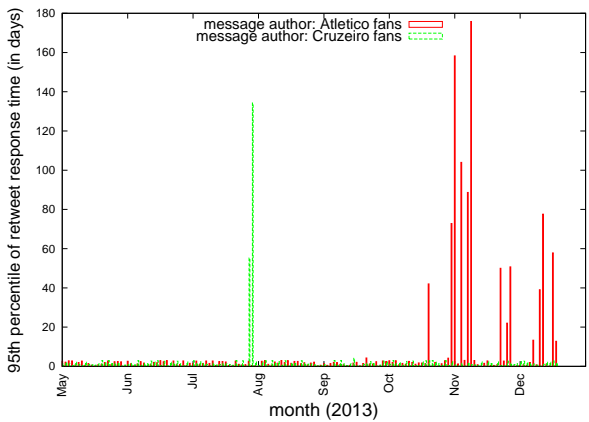}
\caption{95th percentile of retweet response times triggered by each day during 2013. Spikes coincide
with significant real-world events that triggered different reaction on antagonistic groups;
in general, we observe that wrong predictions made by rival communities are retweeted
by rivals when they are proved wrong.}
\label{fig:histograma_2013}
\end{figure}

We group messages at a daily granularity and its source (Atl\'{e}tico or Cruzeiro fans).
For each of these sets, we plot in Figure~\ref{fig:histograma_2013}
the 95th percentile of the retweet reponse times of the messages
posted by each group on that day. We notice that the main events
related to the Brazilian soccer world were captured as spikes:
in July 2013, Atletico won its first Copa Libertadores, what
generated a huge of spike of retweets of Cruzeiro supporters
who tweeted that Atletico would never win the competition.

The remainder of the year was not favorable to Atl\'{e}tico, though.
In November, Cruzeiro won the national league, and in December
Atl\'{e}tico lost the FIFA Club World Cup. The sequence of unfortunate
events for Atl\'{e}tico fans coincides with a series of
spikes in retweets of their old messages by Cruzeiro fans; including
the wrong prediction that 2013 would be a ``great'' (ironically meaning ``bad'')
year for the Cruzeiro club.

As on Finding 3, we also see potential for using the temporal information
associated to retweets to enrich real-time sentiment analysis models, and 
we leave a more thorough exploring of retweets response times in sentiment
analysis algorithms as future work.

\section{Conclusions}
In this paper we explore the observation
that, in the vast majority of social media studies,
especially those based on Facebook and Twitter data, there
is no explicit positive and negative signs encoded
in the edges. Since inferring individual edge polarities in a unsigned graph is not a trivial task, 
most social studies assume that retweets and shares are endorsement interactions.
No specific analysis
on the polarity of the links crossing the communities is usually conducted and antagonism is assumed due to the modular division of the social graphs into two communities historically known to be antagonistic, such as democrats
and republicans.

Although very recent papers on retweeting
activity still qualify retweets as a strictly positive interaction~\cite{BalancingOpposingViews,WhatDoRetweetsIndicate,PredictingIdeologicalFriends},
we show that retweets can actually carry a negative polarity, conveying a sentiment which
is opposite to the one explicited in the tweet's text. We believe 
the neglected impact of negative retweets explain,
in part, the low accuracy levels obtained in some
user polarity classification experiments~\cite{PolarityNotEasy}.
We also demonstrate that negative retweets contribute to make antagonistic
groups \emph{closer} to each other in a network of retweets, what can lead to misleading conclusions
by na\"{\i}ve network models, particularly when multiple communities are found, due to the absence of retweets
between neutral communities.

We found that one of the reasons that motivate Twitter users to broadcast tweets they disagree with
is to create irony by broadcasting a message in a different temporal context, especially when a real-world event that
disproves the original message argument happens. Such behavior finds similarity on \emph{quoting out of context}, 
a practice already described in the Communications literature~\cite{TheyNeverSaidIt}.

We believe the better understanding of retweets as multifaceted social interactions which can be (1) possibly
negative and (2) have a temporal component may support the design of algorithms that exploit the network structure in conjunction with opinionated content to better perform tasks typically offered by social media platforms, such as content recommendation, event detection, sentiment analysis and news curation~\cite{pcalaisKDD11,UserClassification}.

We acknowledge that one of the limitations of our study is that the method that find clusters through
random walks from seed nodes do not distinguish between positive and negative retweets; then, some
users may be wrongly classified exactly due to the ironic broadcasts he may engage in. However, since
positive retweets are still dominant, this effect should affect a few users. Nevertheless, we
can think of algorithms that simultaneously infer both edge polarities and user memberships as an interesting
future work. Another interesting approach would be weighting edges by their retweet response times; community
detection methods could give more priority to recent retweets when seeking for homophilic relationships.

Our work also reinforces the opportunity and possibilities of building rich models which
combine content, network structure and temporal dimensions of the underlying social data. Since
each dimension is ambiguous in nature, powerful predictive and descriptive methods can be built
upon combining these three evidences.

\subsection*{Acknowledgments}

This work was supported by
CNPQ, Fapemig, InWeb, MASWeb, BIGSEA and INCT-MCS.

\begin{quote}
\begin{small}
\bibliographystyle{aaai}
%\bibliography{references}

\end{small}
\end{quote}

\end{document}